%% file: paper8.tex
\documentclass[runningheads,a4paper]{llncs}
\usepackage{eso-pic}
\AddToShipoutPicture*{
\put(133,160){
\scriptsize
Proceedings of WLPE 2013, {\tt arXiv:1308.2055}, August 2013.
}}

\usepackage{llncsdoc}
\usepackage{graphicx}
\usepackage{epsfig}

\usepackage{epstopdf}

\newcommand{\atoms}{atoms}
\newcommand{\mif}{\mbox{ :- }}

\begin{document}
%

%
\title{Understanding Rulelog Computations in Silk}
\authorrunning{Carl Andersen et al.}
\author{
  Carl Andersen\thanks{Raytheon BBN Technologies, USA}
  \and
  Brett Benyo$^*$
  \and
  Miguel Calejo\thanks{Declarativa, Portugal}
  \and
  Mike Dean$^*$
  \and
  Paul Fodor\thanks{Stony Brook University, USA}
  \and
  Benjamin N. Grosof\thanks{
  Benjamin Grosof \& Associates, LLC, USA}
  \and
  Michael Kifer$^{***}$
  \and
  Senlin Liang$^{***}$
  \and
  Terrance Swift\thanks{
  CENTRIA, Universidade Nova de Lisboa, Portugal
    }
\institute{}
}

\maketitle


\begin{abstract} 

Rulelog is a knowledge representation and reasoning language based on
logic programming under the well-founded semantics.  It is an
extension of the language of Flora-2 and so supports inheritance and
other object-oriented features, as well as the higher-order syntax
of Hilog.  However, Rulelog rules may also contain quantifiers and may
be contra-positional.  In addition, these rules are evaluated in the
presence of defeasibility mechanisms that include rule cancellation,
rule priorities, and other aspects.  Rulelog programs are
sometimes developed by loosely coordinated teams of knowledge
engineers (KEs) who are not necessarily programmers.  This
requires not only declarative debugging support, but also support for
{\em profiling} to help KEs understand the overall structure of a
computation, including its termination properties.  The design of
debugging and profiling tools is made more challenging because Rulelog
programs undergo a series of transformations into normal programs, so
that there is a cognitive distance between how rules are specified and
how they are executed.

In this paper, we describe the debugging and profiling environment for
Rulelog implemented in the integrated development environment of the
Silk system.  Our approach includes an interface to justification
graphs, which treat why-not and defeasibility as well as provenance of
the rules supporting answers.  It also includes tools for trace-based
analysis of computations to permit understanding of erroneous
non-termination and of general performance issues. 
For semantically correct cases of the non-terminating behavior,
Silk offers a different approach, which 
addresses the problem
in a formally sound manner by leveraging a form of bounded
rationality called {\em restraint}.



\end{abstract}

\input{intro}

\input{rulelog}

\input{justification}

\input{trace}

\input{disc}


\subsubsection*{Acknowledgements}
This work was supported by Vulcan, Inc., as part of the Halo Advanced
Research project.
Michael Kifer and Senlin Liang were
also partly supported by NSF grant 0964196.
Thanks to Paul Haley (Automata, Inc.)  and Keith
Goolsbey (Cycorp) for helpful discussions, and to Walter Wilson who
implemented {\tt table\_dump}. We also than the anonymous referees for
suggesting the many improvements to an earlier draft.



\bibliography{profiling,all}            
\bibliographystyle{plain}

\end{document}

%% file: intro.tex
\section{Introduction}

Knowledge Representation and Reasoning (KRR) languages are often
expressive fragments of first-order logic, such as OWL-DL \cite{OWL},
or are fixed point logics based on ASP \cite{answer-set-08} and its
extensions.  While such languages offer concise representations and
use powerful reasoning techniques, they often are not scalable enough
to address applications in domains such as the semantic web.  As an
alternative, KRR languages and systems have also been developed whose
inferencing is of lower computational complexity, such as the
description logic ${\cal EL}+$~\cite{BBL05}, or the rule based systems
Flora-2~\cite{YaKZ03} and Silk.

The Silk system is based on a language called Rulelog, which combines a
number of features from logic programming (LP). 
Silk has a 3-layer architecture. The top (Java) layer provides much of the
connectivity to other systems as well as some of the transformations, such
as the \emph{omni} transform discussed later. The second layer, Flora-2,
supplies most of the functionality of Rulelog. The two upper layers compile
Rulelog programs into normal programs (containing recursion and
logical functions) and execute them under the well-founded sematics using
the XSB system~\cite{SwiW12}~\footnote{Rulelog has also been partially
  implemented using the Cyc reasoner of Cycorp.}.  The main
features of Rulelog add considerable expressivity over normal logic
programs.  As with XSB, answers may have an truth value of {\em
  undefined} due to a form of bounded rationality called {\em radial
  restraint}~\cite{GroS13}, in addition to being undefined due to a
loop through negation.  As with Flora-2, the frame-based syntax is
supported, heavy use can be made of Hilog, and rules may be
defeasible.  In addition, Rulelog allows use of a general first-order
syntax not only in rule bodies as in \cite{LlT84}, but also in rule
heads giving rise to {\em omni rules}.
The use of these features raises a number of issues both in debugging
and in understanding the behavior of Rulelog derivations.

Traditionally, the {\em justification problem} is the problem of
explaining missing or unexpected answer~\cite{PGDRR04,PSE09}.
Justification is made more complex by several features of Rulelog.
First, Rulelog inferencing is based on the well-founded semantics,
leading to the task of explaining answers whose truth value is {\em
  undefined}.  In addition, the use of defeasible reasoning and of
Hilog can lead to unexpected inferences.  Furthermore, transformations
are used to implement omni rules, Hilog, and defeasibility.  Together,
these transformations can make the compiled rules look substantially
different than the source.

The {\em performance/termination problem} is the problem of indicating
why a derivation has taken up more resources than expected ---
including non-termination as an extreme case.  Addressing this problem
has been especially important for Rulelog users.  As will be shown
below, new and sometimes cyclic dependencies arise from the use of
omni rules, from defeasibile reasoning, and from the use of Hilog.  While these
cyclic dependencies can be addressed by tabling, unexpected cyclic
dependencies can lead to huge mutually recursive components.  In
addition, as logical functions are allowed, unexpected cycles can lead
to non-terminating queries.  These problems with unexpected dependencies
are exacerbated by  the fact that  Rulelog is aimed at
knowledge engineers (KEs) who are
competent in logic, but who are not necessarily computer programmers.
These KEs share a common background vocabulary in developing
inter-dependent rules, but are often loosely coordinated.

This paper discusses the approaches to the justification and
performance problems that are part of the Silk system.
Section~\ref{sec:rulelog} discusses the Rulelog language and how it
has been implemented via Flora-2 and XSB.
Section~\ref{sec:justification} discusses the approaches to the
justification problem, while Section~\ref{sec:trace} discusses how
performance is assessed via traces of Flora-2 and XSB query
evaluations.  We note that many of these approaches are still under
development, and in each section we note the current status of each
approach.


%% file: rulelog.tex
\section{Rulelog and its Implementation} \label{sec:rulelog}

Rulelog, as implemented in the Silk system, supports not only direct
construction of knowledge bases, but also interchange with knowledge
formatted in various semantic standards such as OWL-RL and RDF.
Rulelog also provides a good target for text-based
authoring of knowledge \cite{grosof:tleo-ruleml-2013}, because of its
ability to express defeasible formulas as axioms.  In this section we
present the main semantic and syntactic features of Rulelog that
affect debugging and profiling, and then briefly discuss its
implementation in the Silk system.


\subsection{Omni Rules and Hilog} \label{sec:omni}
In this paper, we make use of general terminology of logic programs,
but adopt some syntax of Rulelog.  In Rulelog, varibles are designated
by tokens that begin with $'?'$; a {\em literal} is an atom $A$ or its
explicit negation $neg\ A$; a {\em default literal} is either an
objective literal $O$ or its default negation {\em naf O}.\footnote{In
  the literature, literals in our terminology are sometimes called
  objective literals, and default literals are called literals.}  A
normal rule is designated as
$
   Head \mif{} L_1\ and \ldots and\ L_n,
$
where $Head$ is an atom and each $L_i$ is a default literal.
Bodies of normal rules can be made to have a more expressive syntactic
form using the well-known Lloyd-Topor (LT) transformation\cite{LlT84}.
For example, the rule
\begin{tabbing}
foo\=foo\=\kill
\> {\em      p\_equivalent(?X,?Y) :- forall(?Z)(p(?Z,?X) $<==>$ p(?Z,?Y)).}
\end{tabbing}
is LT-transformed into a sequence of normal rules.

{\em Omni rules} extend the LT-transformation by allowing first-order
formulas to occur in rule heads as well as rule bodies.  To take an
example, a statement of molecular biology {\em ``A contractile vacuole
  is inactive in an isotonic environment''} can be modeled using the
omni rule
{\small
\begin{tabbing}
foo\=foo\=\kill
{\em forall(?x6)\^{}contractile(vacuole)(?x6)) } \\
\> {\em     ==$>$ forall(?x9)\^{}isotonic(environment)(?x9) } \\
\> {\em     ==$>$ inactive(in(?x9))(?x6). }
\end{tabbing}
which is transformed into the rules
\begin{tabbing}
foooo\=foo\=\kill
{\em neg isotonic(environment)(?x9) :- } \\
\> {\em        contractile(vacuole)(?x6) and neg inactive(in(?x9))(?x6). }
\vspace{2mm}\\
{\em neg contractile(vacuole)(?x6) :- }\\
\> {\em        isotonic(environment)(?x9) and neg inactive(in(?x9))(?x6).}
\vspace{2mm}\\
{\em inactive(in(?x9))(?x6) :- }\\
\> {\em        contractile(vacuole)(?x6) and isotonic(environment)(?x9).}
\end{tabbing}
}

As can be seen from this example, the semantics of omni rules is
entirely transformational.  Note that each of these rules makes use of
Hilog, which allows any functor of a term to be a variable or a
compound term, rather than simply an atom.  Flora-2 itself allows
explicit negation, and applies another transformation to remove the
explicit negation, resulting in normal rules executed by XSB.

\subsection{Rule Descriptors} \label{sec:rule-descriptors}

While Prolog or ASP programmers typically think in terms of
predicates, KEs often think in terms of rules when they want to
understand inferences or make declarations about overriding.
Accordingly, Rulelog supports meta-information about rules through
{\em descriptors}, which rely on the frame syntax of Flora-2 and may
themselves be defined as rules.  As an example, consider a rule
indicating that a eukaryotic cell has a visible nucleus as in the
following code.
\begin{tabbing}
foo\=foo\=\kill

{\em @!\{r1[tag\texttt{->}eukc, kgroup\texttt{->}g1]\}}  \\
\> {\em  eukaryotic(cell)(?x1) :- has(?x1,?x2) and visible(nucleus)(?x2).} \\
\\   
{\em @@strict   ?x[owned\texttt{->}Benj] :- @!\{?x[kgroup \texttt{->}g1]\}.}
\end{tabbing}
The above syntax means that the rule about eukaryotic
cells has the unique identifier {\em r1}; the rule also has a {\em tag}
attribute whose value is {\em eukc} (rule tags are used by argumentation
theories to determine whether the rule can be opposed or
overridden).
The second rule has a descriptor \texttt{@@strict}, which means that the
rule is not defeasible. The rule itself says that any rule whose descriptor
meta-information says that it belongs to the \emph{kgroup} \emph{g1} is
owned by \emph{Benj}.   

\subsection{Argumentation Theories} \label{sec:defeasible} 
As a KRR language, Silk makes heavy use of argumentation theories
\cite{WGKFL09} that
affect the derivations made by rules.
Defeasibility is specified via two
user-defined predicates: {\em opposes/2} and {\em overrides/2}.  Two
atomic formulas {\em oppose} each other if no model of a program may contain
both atoms: e.g., an atom and its explicit negation oppose each other,
but opposition can capture other types of contradictions.
In addition, rules can be prioritized. Each rule has an explicit or
implicit \emph{tag}. Implicit tags default to rule ids, but tags generally
are distinct from rule ids: tags are used for prioritizing rules, so 
different rules may have the same tag. 
Given two rules, one may {\em override} the other and so be given
preference, which is written as \emph{override(tag1,tag2)}, where
\emph{tag1} and \emph{tag2} are the tags of the respective rules.
Figure~\ref{fig:courteous} shows a highly simplified argumentation
theory.  Such a theory is applied to a
program by a transformation as follows~\cite{WGKFL09}.  Each
clause $@\{id[tag\texttt{->}t]\} H \mif{} B$ whose head is a defeasible predicate is rewritten
as $H \mif{} B, naf\ defeated(t)$; clauses for non-defeasible rules (called
{\em strict}) are not altered.  For atoms $A_1$ and $A_2$,
if $A_1$ and $A_2$ are both derivable and oppose each other but
neither overrides the other, $A_1$ and $A_2$ mutually {\em rebut} each
other.  If, in addition, $A_1$ overrides $A_2$, we say that $A_1$ {\em refutes}
$A_2$.  Within Silk, the compilation of a argumentation
theory ensures that rebutted atoms have an {\em undefined} truth
value, as do atoms that refute themselves (i.e. if the {\em
  overrides/2} predicate is cyclic).

\begin{figure}
{\em
\small{
\begin{tabbing}
foofoooooooooooooooooooooooooooooooooooo\=\kill
defeated(?T) :- (refutes(?T2,?T) or rebuts(?T2,?T)), candidate(?T2). \\
refutes(?T,?T2) :- conflicts(?T,?T2), overrides(?T,?T2). \\
rebuts(?T,?T2) :- conflicts(?T,?T2), naf overrides(?T,?T2). \\
conflicts(?T,?T2) :- (opposes(?T,?T2) or opposes(?T2,?T)),candidate(?T2). \\
candidate(?T) :- headof(?T,?H), ?H.
\end{tabbing}
} }
\centering
\caption{A Highly Simplified Defeasibility Theory}
\label{fig:courteous}
\end{figure}
Argumentation theories used in practice are far more complex than that
of Figure~\ref{fig:courteous}.  A primary motivation is to capture
sophisticated arguments a human reasoner might make in applying default
reasoning to a problem.
In justifying Rulelog inferences, it is important to communicate to the
user the steps when a truth value
is concluded due to the use of defeasibility.

\subsection{Use of Three-Valued Logic} \label{subsec:restraint}

The well-founded semantics assigns a partial model to a program $P$,
where the truth value of certain atoms in the ground instantiation of
$P$ may be neither true nor false, but {\em undefined}.  This third
truth value was added to handle situations in which an atom $A$ had no
true derivations; but where $A$ had at least one derivation which
depended on $naf\ A$, a situation we term a {\em negative loop}.

In systems like XSB and Flora-2, which support the well-founded
semantics, the third truth value can be put to use for other reasons
as well.  For instance, in ISO Prolog exception (or error) conditions
cause a computation to abort if the error is not specifically caught.
However, this procedural approach to handling errors could be replaced
by a semantics that assigns an atom $A$ a truth value of {\em
  undefined} when $A$ has no true derivations, but $A$ has at least
one derivation that depends on an exception.  This approach properly
generalizes the well-founded semantics from negative loops to other
classes of exceptions.  As noted in~\cite{Nais06}, a three-valued
approach can be superior for debugging to the ISO Prolog approach, as
the entire search space for proving an atom $A$ can be examined,
including possible true derivations of $A$ that might not be seen if
an exception were thrown.

\paragraph*{Restraint} 
The third truth value may be generalized yet again, leading to a type
of bounded rationality termed {\em restraint}.  Consider the program:

{\small  {\em 
\begin{tabbing}
fooooofooooofooooo\=fooooofooooofooooofooooo\=foo\=\kill
p(0). \> p(s(X)):- p(X). \> p(X):- p(s(X)). 
\end{tabbing}
} }

\noindent
If the goal {\em ?- p(a)} is given to this program, then a tabled
evaluation may create an infinite number of subgoals: {\em p(a)},{\em
  p(s(a))},{\em p(s(s(a)))}, and so on. In addition, the goal {\em
  p(Y)} to the above program has an infinite number of answers.

Each of these situations can each be handled by a different tabling technique.
If {\em subgoal abstraction} is used and if a program is {\em safe}
(as is the preceding program), then it can be ensured that only a
finite number of subgoals are created~\cite{RigS13a}.\footnote{Note
  that the tabling technique of call subsumption will not help as none
  of the above subgoals subsume one another.}  If the technique of
{\em radial restraint}~\cite{GroS13} is used, it can be ensured that
only a finite number of answers are generated.
To see how this works, suppose that the following declaration were
made for $p/1$:

{\small  {\em 
\begin{tabbing}
foooooooooooooooooooooooo\=foo\=\kill
:- table p/1 as subgoal\_abstract(2),answer\_abstract(3).
\end{tabbing}
} }

\noindent
Such a specification causes abstraction in two cases.  Any subterm in
a subgoal for this predicate whose depth is greater than 2 is
abstracted (replaced with a variable), so that the only subgoals
created for {\em ?- p(a)} would be {\em p(a)} and {\em p(s(X))}.
Similarly, any subterm in an answer for this predicate whose depth is
greater than 3 would also be abstracted {\em and assigned the truth
  value {\em undefined}}.  Thus, the goal {\em ?- p(Y)} would return
the answers {\em p(0)}, {\em p(s(0))} and {\em p(s(s(0)))}; but {\em
  p((s(s(s(0)))))}, {\em p((s(s(s(s(0))))))}, etc. would be abstracted
to {\em p((s(s(s(X))))} (and assigned the truth value of {\em
  undefined}.  Because radial restraint uses the truth value {\em
  undefined} to answers when abstracting, it preserves the soundness
of derivations, even in the presence of negation.

As a form of {\em unsafety restraint}, the XSB predicate {\em unot/1}
is a type of tabled negation that makes the default literal {\em
  unot(A)} undefined if {\em A} is non-ground.  This can be seen as a
type of restraint, as in principle constructive negation could be
used, though constructive negation might raise the computational
complexity of a derivation, or make a derivation non-terminating.  In
Flora-2, the use of {\em unot/1} is combined with the ability to delay
evaluation of non-ground negative subgoals through the use of the {\em
  when} library.  The truth value of a non-ground negative subgoal is
made {\em undefined} only when it is determined that variables in the
subgoal can never be instantiated, so that further delaying is futile.

Using {\em rule-based} restraint, general forms of bounded rationality
can be programmed explicitly.  Conceptually, rule-based restraint is
invoked using rules for {\em skipping} similar to those for {\em
  opposes} and {\em overrides} (Section~\ref{sec:defeasible}).
Skipping is easiest understand by example.  Consider the following fact and
a rule:

{\small  {\em 
\begin{tabbing}
foo\=fooooooooooooooooo\=foo\=\kill
\>step(1). \> step(N\_out):- step(N),N\_out is N + 1.
\end{tabbing}
} }

\noindent
which might be used, e.g., to define a planning horizon.  Note that
radial restraint could not be used to soundly restrict the planning
horizon generated by {\em step/1} as integer arithmetic does not use
the term structure needed by that technique.  However the rule:

{\small  {\em 
\begin{tabbing}
foo\=foooooooooooooooooooooooo\=foo\=\kill
\>skip(step(N),[N],[\_]):- N $>$ 10.
\end{tabbing}
} }
\noindent
would cause the body of the second rule for {\em step/1} to be compiled to 
{\small  {\em 
\begin{tabbing}
foo\=foooooooooooo\=foo\=\kill
\>step(N\_out):- step(N),N1 is N + 1, \\
\>\> (skip(step(N1)) $\rightarrow$ N\_out = N1 ; undefined). 
\end{tabbing}
} }
\noindent
If the body of the skip statement is false ($N =< 10$), no rewrite is
made to the variable, {\em N\_out} of $step/1$.  Otherwise {\em
  N\_out} is abstracted to an anonymous variable, and the goal {\em
  undefined} is called, so that {\em step(N)} is {\em undefined} when
$N > 10$.  Thus the goal {\em step(X)} would have 11 answers, where
the first 10 bind {\em X} to the intergers 1 to 10 with truth value {\em
  true}, while the eleventh does not bind {\em X} and has the truth
value {\em undefined}.

Note that a similar skipping rule could be used to program both radial
and unsafety restraint, so that rule-based restraint is more general
than either of these approaches.  However, both radial restraint and
unsafety restraint are implemented in the XSB engine, so are more
efficient than rule-based restraint.  Rule-based restraint can be
mixed into an argumentation theory, giving rise to {\em restrained
  argumentation theories}.




%% file: justification.tex
\section{Justification}  \label{sec:justification}

\emph{Justification}, a method of explaining how a derivation was
made, has a long history in KRR, starting with \emph{truth maintenance
  systems} (e.g., \cite{mcallester90truth}) and applied in LP to
tabled logic programming~\cite{PGDRR04} and ASP~\cite{PSE09}.
Justification for Rulelog requires addressing the various features
described in Section~\ref{sec:rulelog}.  For instance, recall that,
in defeasible reasoning, a literal might be {\em false}
or {\em undefined} because it is derived by rules that are
\emph{defeated} by other rules. In those cases, it is necessary to
explain how and why those rules were defeated, and whether
prioritization was involved.  A key aspect is not only to explain why
literals are {\em true} or rules are active, but why literals are {\em
  false} (or {\em undefined}), and rules are defeated.  In addition,
if a literal has a truth value {\em undefined}, the user should be
able to understand whether this is due to a negative loop, to
restraint, to unsound negation, or some other error condition---or due to
some combination of theses reasons.  In addition, if the inheritance
mechanisms of Flora-2 are used, an explanation must be given to why a
given predicate was chosen to derive an attribute value for an object.

\begin{figure*}[t]
\begin{center}
\includegraphics[width=13cm]{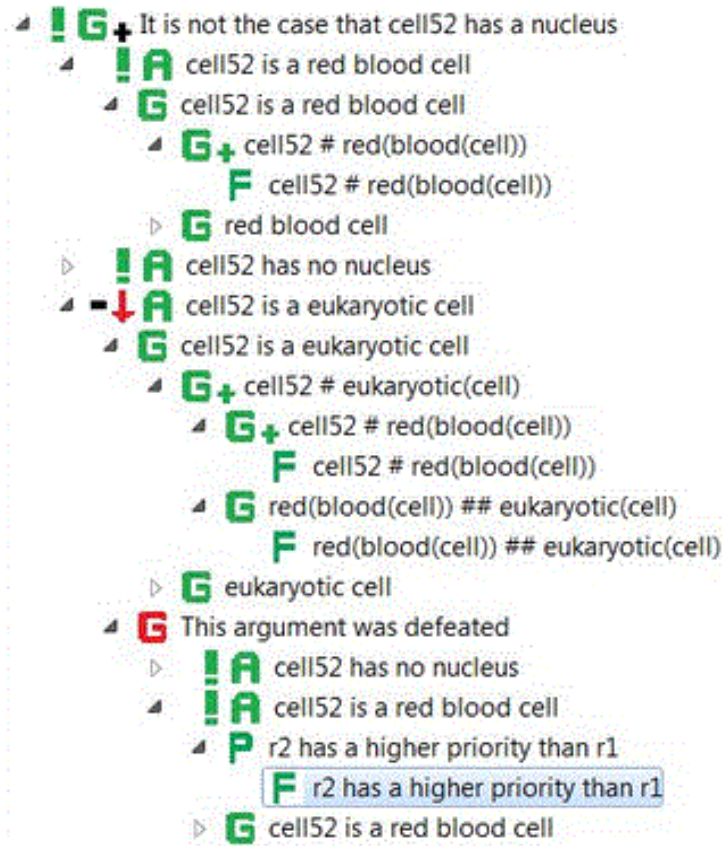}
\end{center}
\caption{Justification example} 
\label{fig:jg2}
\end{figure*}

Within Silk, justification is visualized graphically, through the
Silklipse environment (\cite{silk-justif-ruleml2010} described an
early version).  In addition, Silk supports rule-based transformation
of the justification information: allowing literals to be displayed as
English text, permitting sets of literals to be summarized or omitted
from the justification, and so on.
Figure~\ref{fig:jg2} shows a screenshot of a navigable justification
in Silklipse for why the statement ``{\em cell53} has a nucleus''
is (default) {\em false}.  Justification rules have transformed some
lines into English text; for instance, the first line reads: {\em ``It
  is not the case that cell52 has a nucleus.''}.  Other lines appear
as Rulelog literals, such as the forth line:
{\em cell52 \# red(blood(cell))),}
where {\em ``\#''} means {\em ``is an instance of the class''}.  Each
line is associated with one or more icons that indicate how a literal
has been used in a derivation.  The icon {\em ``G''} indicates a
literal (perhaps translated into English) that is a (sub)goal.  {\em
  ``A''} indicates an argument for support of the literal--- in the
sense of prioritized argumentation in defeasibility.  Operationally an
argument may be a rule body, but can also include literals from the
head of a contra-positional omni rule.  {\em ``F''} indicates a fact,
i.e., a literal that was directly asserted, while {\em ``P''}
indicates prioritization meta-information, i.e., that one rule's tag
overrides another tag.  The color green indicates that a literal is
true, while red indicates that a literal is false.  Similarly, a green
bang (``!'')  indicates a undefeated argument, while a red down arrow
(``$\downarrow$'') indicates an argument that has been defeated---perhaps refuted by conflicting arguments with higher priority.  The
plus sign (``$+$'') just to the right of {\em ``G''} indicates that
there are more arguments to see.  When the ``$+$'' is black it
indicates there are both pro and con arguments to see; when green, it
indicates there are more pro arguments but not more con arguments to
see.  Finally, a black bar (``---'') indicates an argument for the
{\em neg} (strong negation) of the goal literal.

In this example, the relevant asserted logical rules in the KB can be described
in English as follows:   
{\em 
\begin{quote}
cell52 is a red blood cell. \\
Eukaryotic cells have nuclei.  (This rule has tag r1.)   \\
Red blood cells are a subclass of eukaryotic cells.\\
Red blood cells do not have nuclei.  (This rule has tag r2.) \\
r2 has higher priority than r1.  
\end{quote}}

Silk's justification system has been heavily used by KEs, so that
obtaining good performance for the justification has been a critical
task.  The first approch was to make use of a justification
transformation of the rules, in a spirit similar to~\cite{PGDRR04}.
This approach quickly proved unwieldy due to the transformations
needed for defeasibility and for omni rules.  For instance,
the omni rule in Section~\ref{sec:omni} generated 98 justification
rules.  As an alternative, Silk now invokes a meta-interpreter to
construct a justification graph when a user requests justification.
The heavy use of tabling in Flora-2 means that the justification graph
can be reconstructed quickly in most cases by making use of
information in the tables themselves.

\paragraph*{Status of Justification}
The current version of justiification in Silk was tested on a previous
generation of argumentation theories, and does not yet fully support
the deeper argument refutations used in ATCO
(Section~\ref{sec:defeasible}), or the various types of restraint
discussed in Section~\ref{subsec:restraint}.  However, even with these
limitations, it has been actively used by KEs.


%% file: trace.tex
\section{Performance Profiling through Trace-based Analysis} \label{sec:trace}

Justification as previously described addresses why a literal is {\em
  true}, {\em false} or {\em undefined}, and so relies on the
semantics of Rulelog.  Accordingly, the justification graph itself is
largely independent both of the transformations of Rulelog into
Flora-2 code and then into normal programs, and of the tabled
evaluation used by XSB.  Understanding the performance profile of
derivation (and to some extent whether it terminates) depends on the
characteristics of how a derivation was actually implemented.
Performance bottlenecks will differ depending on whether a derivation
is based on tabling or on bottom-up evaluation; on whether the tabling
uses call-variance or call subsumption~\cite{SwiW12}, and so on.
The tools presented here for trace-based analysis rely on
characteristics of XSB's tabled evaluation, but the user-level tools
try to minimize the amount of information a user needs to know about
the particulars of tabling.

\subsection{Table Dump:  Examining Subqueries, Answers, and Rules}

\begin{figure*}[tb]
\begin{center}
\includegraphics[width=13cm]{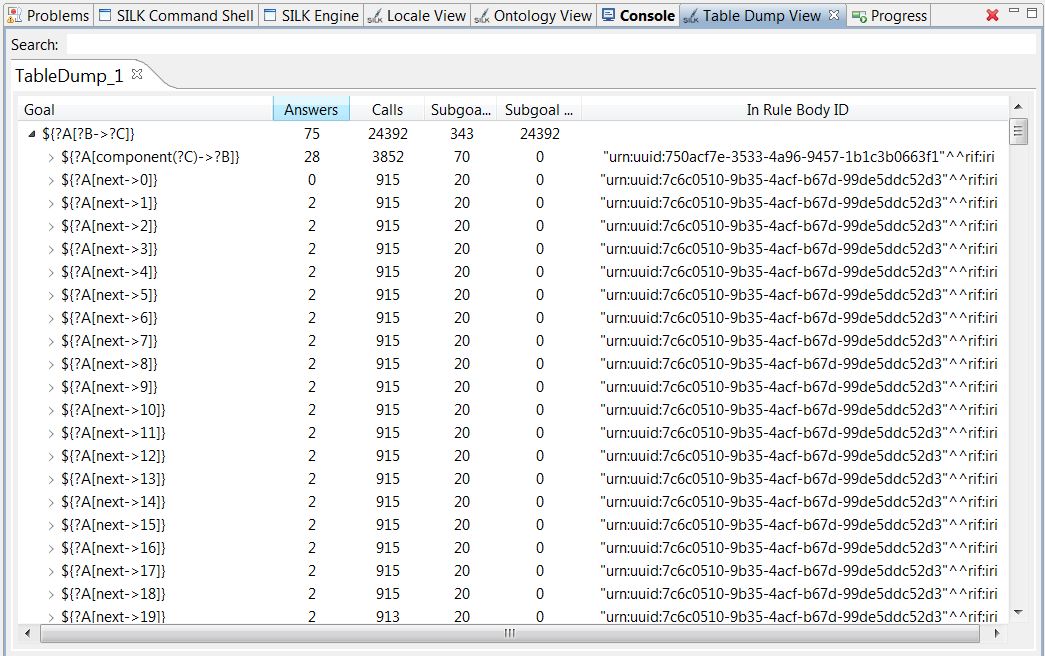} 
\end{center}
\caption{Table Dump example}
\label{fig:td4}
\end{figure*}

The most direct way to understand the performance of a tabled
computation is to look at the tables themselves.  \emph{Table Dump} is a
tool that does just that.  A user enters a term $T$, and Table Dump
returns information about all tables whose subgoals are subsumed by
$T$.  If $T$ is variable, information about all tables is returned.
As Silk computations may produce hundreds of thousands of tables,
information can be returned in a summary form, and users may
``drill-down'' by querying successively more instantiated terms.
Figure~\ref{fig:td4} shows a screenshot of a navigable view of table
dump information in Silklipse for the schematic Rulelog term {\em
  ?A[?B-$>$?C]} --- a frame syntax for asking about
attribute-value pairs {\em [?B-$>$?C]} for some object {\em ?A}.
Information is given about the answers to a subgoal, the number of
times it has been called, and even the rule (or rules) used to call
the subgoal.  Table Dump thus helps a KE to identify bottlenecks in
the knowledge base and then take measures such as adding appropriate
guards to rules or reordering subgoals within rules.

The Table Dump tool for Silk is based in part on the {\tt table\_dump}
family of predicates in XSB.  XSB's {\tt table\_dump} is implemented
using XSB's table inspection predicates, and relies on the XSB engine's
maintenance of a count of the number of times each table is called.
However XSB's tables maintain information about subgoals, and does not
provide the rule-specific information seen in Figure~\ref{fig:td4}.
To support this, Flora-2 performs a transformation
that embeds rule ids in each predicate of the rules in question.
These predicates are tabled so that when information is returned from XSB's
{\tt table\_dump}, information about the rules is extracted and shown
  in Silklipse.

\subsection{Trace-Based Analysis Based on Forest Logging}

Although simple and powerful, the table dump approach lacks two main
features needed to fully address the performance/termination problem.
First, it does not provide an overview of how given subqueries in a
derivation relate to one another through rules, and does not display
information about the recursive components whose computation is
central to a Rulelog derivation.  Second, no information is provided
about the order of events in a derivation, such as when subqueries
were made, answers derived, and so on.  

Within Silk, details of a Rulelog derivation can be reconstructed
through another kind of trace-based analysis.  
XSB provides a mechanism to create a more dynamic 
trace or log of a derivation, called a {\em forest log} \cite{Swif12a}.  
Using such a 
log, the structure of even very large recursive components can be
analyzed, and non-terminating derivations detected.  This subsection
first overviews forest logs and afterwards discusses the analysis
routines based on the logs.

\subsubsection{Forest Logging}
The form of tabling used by XSB is called SLG resolution.  The
operational semantics of SLG evaluation (and hence a Rulelog
derivation) can be modeled as a sequence of forests of trees, where
each tree corresponds to a tabled subquery $S$ and represents the
immediate subqueries that $S$ produces along with any answers to $S$.
In fact, each SLG operation is modeled as a function from forests to
forests that creates a new tree, or adds a node or label to an
existing tree.

Within XSB, SLG resolution is executed using a byte-code virtual
machine analogous to that used by Java.  
An internal XSB
flag can be set so that any byte-code instruction that
corresponds to a tabling operation will log information about itself
and its operands as a Prolog-readable term.  For instance, if (tabled)
subgoal $S_1$ is called in the context of subgoal $S_2$, and it is the
first time $S_1$ is called in an evaluation, a fact of the form 

{\em table\_call($S_1$,$S_2$,new,ctr)}

\noindent
is logged, where {\em ctr} (mnemonic for ``counter'') 
is a sequence number for the fact.  When a
derivation ends or is interrupted, the log can be loaded into XSB and
analyzed as a set of Prolog facts.  Within XSB, the logging system is
written at a very low level for efficiency.  Turning on full logging
usually does not slow down performance by more than 70-80\%.
XSB also provides routines to load logs and index their facts on
various arguments.  Based on the logging libraries, logs containing
hundreds of millions of facts have been loaded and analyzed~\cite{Swif12a}.

\subsubsection{Analyzing Recursive Components} \label{sec:sccs}

Once a log has been loaded, a user may ask for an overview of a
computation, which provides information on the total number of calls
to tabled subgoals, the number of distinct tabled subgoals, the number
of answers, and so on.  In addition, the overview provides aggregate
information on the number of mutually recursive components, and the
number of subgoals in the components.  Finally, the overview contains
information indicating how stratified the negation was in a derivation
by listing the total number of \atoms{} whose truth value was
undefined, along with a count of the various SLG operations used to
evaluate well-founded negation.

Some derivations may give rise to very large recursive
components---due to an unanticipated effects of the Hilog, omni, and
defeasibility transformations; to a knowledge base that is not
sufficiently modularized, or to other reasons.  The analysis routines
allow a given recursive component to be examined, by listing the
subqueries in the component, along with the pairs of calling and
called subgoals within the component.

By examining this output, users can usually fix whatever problems gave
rise to large recursive components.  However for a very large
component ${\cal C}$, the number of subqueries in ${\cal C}$ may be on
the order of $10^5$ or more and the number of calling/caller pairs may
be on the order of $10^6$.  In such a case, displaying every subquery or
pair may be confusing at best.  The analysis routines thus provide
several abstraction routines that allow a user to coalesce similar
\atoms .  For instance, if a component contained the subqueries {\em
  p(a,X)}, {\em p(b,X)}, {\em p(c,X)} ..., the analysis routines could
use {\em mode abstraction} to coalesce all of these terms to {\em
  p(bound,free)}, or even {\em predicate abstraction} to coalesce all
these terms to {\em p/2}.  Recursive component analysis together with
abstraction of \atoms{} has been used to analyze the behavior of
reasoning that was translated from Cycorp's inference engine into the
Silk implementation of Rulelog, among other applications.


\subsubsection{Analyzing Runaway: Terminyzer}

Runaway computation occurs when a query does not terminate or takes
too long to come back with an answer. The first type of problem occurs
typically due to the presence of function symbols and the second is
largely due to computations that produce very large intermediate
results most of which could be avoided with smarter evaluation
strategies, such as subgoal reordering.

One sophisticated diagnostic tool we have developed to tackle the
non-termination problem is called the \emph{Terminyzer} (short for
``(non-)Termination Analyzer'') 
\cite{LiaK13a}.  This tool relies on the previously described
\emph{forest logging} mechanism, which records the various tabling
events that occur in the underlying inference engine XSB
\cite{SwiW12}. Among others, forest logging records when the different
subgoals are called and when they receive answers.  Terminyzer
performs different kinds of analysis, such as \emph{call-sequence
  analysis} and \emph{answer-flow analysis}, and identifies the
sequences of subgoals and rules that are being repeatedly called and
in this way cause non-terminating computation.

Terminyzer also has a heuristic that can suggest to the user that the
system be allowed
to reorder subgoals at run time and thus avoid non-termination. For
instance, in a composite subgoal \emph{p(?X,?Y), q(?X)} 
Terminyzer may detect that \emph{p(?X,?Y)} is an infinite
predicate. However, this infinity may be due to the infinite number of
\emph{?X}-values. If \emph{q(?X)} is finite and is evaluated first,
non-termination will not occur. In such a case, Terminyzer may
suggest the user to wrap the offending subgoal with a suitable delay
quantifier---a novel facility supported by Flora-2 and Silk. For instance,
if the above subgoal is rewritten as
\emph{wish(ground(?X))\^{}p(?X,?Y), q(?X)}, the system will not try to
evaluate \emph{p(?X,?Y)} unless \emph{?X} is bound. If it is not bound,
the evaluation of \emph{p(?X,?Y)} is postponed and \emph{q(?X)} will be
evaluated next. If this binds \emph{?X} then all is well and
\emph{p(?X,?Y)} can be evaluated next without a runaway. If \emph{?X}
is still unbound, some other subgoal may possibly bind it, so
\emph{p(?X,?Y)} remains delayed. Only when the system determines that
\emph{?X} cannot be bound no matter what, is \emph{p(?X,?Y)} submitted
for evaluation. If this happens, the user would have to use the information
provided by Terminyzer to decide whether the runaway is a mistake or is
semantically justified. In the first case, this information will help the
user fix the mistake; in the second, restraint (discussed in
Section~\ref{subsec:restraint})
could be used to prevent the runaway.


\subsubsection{Invoking Trace-Based Analysis}
The interfaces to trace-based analysis are based on an XSB 
meta-predicate, called {\tt timed\_call(Goal,Handler,Interval)}, which
calls {\em Goal} and interrupts the computation of {\em Goal} every
{\em Interval} milliseconds to call {\em Handler}.
When {\em Handler}
terminates, {\em Goal} is resumed.  Silk's use of Interprolog allows
XSB and Java processes to call each other recursively.  Thus in
Silklipse, the graphical interface to Silk, {\em Handler} is a call to an interrupt window that allows
access to Table Dump, forest logging, Terminyzer, and other routines.
Given this facility a user may specify that a computation that is
expected to be long running be interrupted every $N$ seconds.  At
interrupt time, the user may explore the current tables in the
computation, turn forest logging on or off, and perform termination
analysis.  Depending on the information given, the user may then
either continue or abort the computation.

\begin{figure}
\includegraphics[width=13cm]{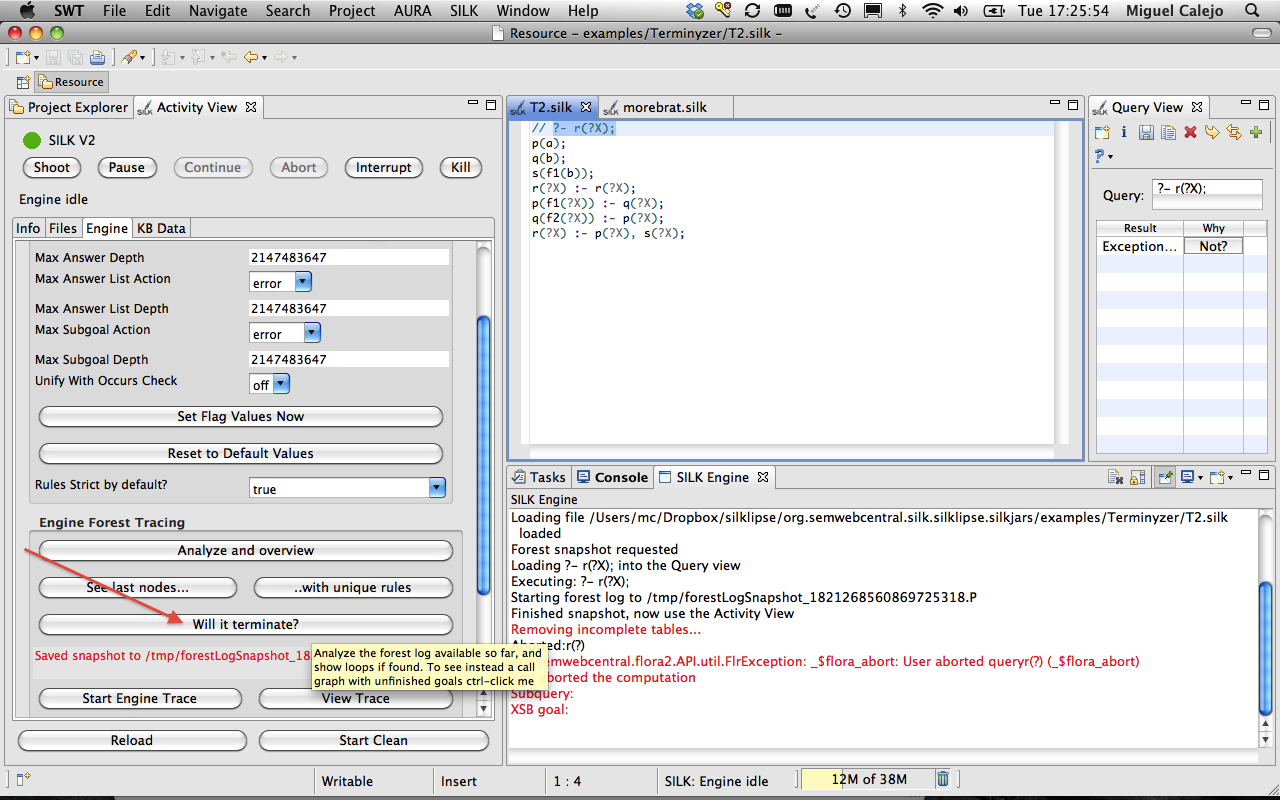} 
\caption{Choices for Trace-Based Analysis at Interrupt Time}\label{fig:terminyzer}
\end{figure}

\paragraph*{Status of Trace-Based Analysis.}
The Table Dump tool and forest logging are both fully integrated into
the Silklipse environment and interrupt mechanism, as are hooks for
Terminyzer (Figure~\ref{fig:terminyzer} shows an example).  The tools in
Section~\ref{sec:sccs} for analyzing recursive components are not yet
available from Silklipse.

%% file: disc.tex
\section{Discussion}

As the Rulelog engine was implemented in Flora-2 and XSB, features of
these systems sparked further development of Rulelog, in particular
the development of restraint.  Typically, the development of debugging
and profiling facilities and their inclusion into Silklipse has
trailed engine development.  Accordingly Silk's justification system
does not yet fully support the features of ATCO or restraint and
analysis of mutually recursive components is not yet implemented
within the Silklipse interrupt screen.  All in all, the development of
Rulelog and Silk have spurred research into extensions of previously
known debugging techniques, such as justification.  They have also
fostered the new trace-based analysis techniques of forest logging and
the Terminyzer, along with ongoing research into better explanations
of the use of restraint within computations.

Funding for Silk from Vulcan Inc. terminated in April 2013.  Vulcan
agreed to allow the Flora-2 code to become open-source, so that all of
the Rulelog engine, with the exception of omni rules, is now fully
open-source.\footnote{All Vulcan-funded work on XSB has been
  open-source.}  Although Vulcan retains rights to the Silklipse
code, a new development environment for Rulelog, called {\em Fidji} is
under active development by the authors of this paper, with the goal
of a prototype release by the end of 2013.




